\documentstyle[12pt,epsf]{article}
\newcommand{\nobody}{\rule{0ex}{1ex}}

\newcommand{\vc}{\char'24\hspace{-1ex}c}
\begin{document}
\voffset -2cm
\title{
\vspace{-2.5cm}
\begin{flushleft}
{\large DESY 91-154}
\end{flushleft}
\vspace{1.5cm} {\LARGE
 Model Independent $Z'$ Constraints\\ at Future $e^+e^-$ Colliders}
}
\author{A. Leike\\
DESY - Institut f\"ur Hochenergiephysik,\\ Platanenallee 6, 15735 Zeuthen,
Germany}
\maketitle
\begin{abstract}
Model independent constraints on the mass of extra neutral gauge bosons and
their couplings to charged leptons are given for LEP~II and a 500\,GeV $e^+e^-$
collider.  Analytical exclusion limits are derived in the Born approximation.
The $Z'$ limits obtained with radiative corrections are always worse than
those calculated at the Born level. Polarized beams are only useful for
degrees of polarization essentially larger than 50\%. Known discovery limits
on extra $Z$ bosons predicted by popular $Z'$ models are reproduced as
special cases.  The $Z'$ constraints are compared to those predicted by four
fermion contact interactions. \vspace{1cm}
\end{abstract}
%
\section{Introduction}
The Standard Model (SM) gives a very successful description of the present
high energy physics data. Its predictions are proven at  the LEP~I energy
range with high precision being sensitive to physics at the one loop level.
However, there is a common belief that the strong and electroweak
interactions of the SM should have a common origin. Their unification is
often done in a large gauge group at higher energies.  The embedding of the
SM in a gauge group larger than SU(5) typically predicts extra neutral gauge
bosons denoted by a $Z'$ here. The search for such a particle is an important
task of any present and future collider.

Up to now no extra neutral gauge boson is found.  The consistency of
experimental data with the SM is usually interpreted in terms of
$Z'$ mass limits for specific $Z'$ models as, for
example,  the $E_6$ GUT \cite{e6,e6lr}, the Left Right Symmetric Model (LRM)
\cite{e6lr,lr} or the Sequential Standard Model (SSM).
Although these limits give an important feeling about the predictive power of
an experiment, the full information about a $Z'$ can
only be obtained by a model independent $Z'$ analysis of the experimental data.
Such a general analysis is possible for the leptonic couplings of the $Z'$
due to the large number of pure leptonic observables at $e^+e^-$ colliders.
The hadronic observables are sensitive to the leptonic couplings of the
$Z'$ too. Therefore, the measurement of the $Z'$ couplings to
quarks in $e^+e^-$ collisions makes sense only, if non-zero $Z'$ couplings to
charged leptons are found.

At hadron colliders, the squares of the $Z'$ couplings to quarks and leptons
can be measured \cite{zphad}. Although the sign of the couplings cannot be
defined there, even a $Z'$ with zero couplings to charged leptons could be
detected via the search for rare processes. As we will show later,  $e^+e^-$
colliders allow a discrimination between different signs of $Z'$ couplings.
Hadron and $e^+e^-$ colliders are complementary in $Z'$ search.

In this paper, we make a model independent $Z'$  analysis and show, how the
different leptonic observables compete in constraining the leptonic $Z'$
couplings.
Known limits for specific $Z'$ models and for four fermion contact
interactions are reproduced as special cases. At the
Born level, approximate analytical formulae are obtained. They make the
analysis and its dependence on the experimental errors transparent. They are
not very different from the final result which contains
all needed radiative corrections. Radiative corrections have to be included
because a $Z'$ gives no events with special signature but shows up by
(probably small) deviations of observables from their SM predictions.
\vspace{1ex}

Taking into account present experimental limits on the $Z'$
mass \cite{cdfzp} and the expected improvements in near future \cite{hewett},
LEP~II and the Next Linear $e^+e^-$ Collider with a c.m. energy of 500\,GeV
(NLC) will probably operate well below a $Z'$ peak.
Further, the $ZZ'$ mixing angle is known to be small \cite{zmix} and can be
set to zero in our analysis.

Larger gauge groups predict not only extra gauge bosons but also additional
fermions. Their effects are described, for example, in \cite{e6,maa} and will
not be considered here.
\vspace{1ex}

In section~2, we give the basic notations of a model independent $Z'$
description and show the connections to some popular $Z'$ models and to four
fermion contact interactions.
Section~3 contains the $Z'$
analysis at the Born level leading to approximate analytical formulae.
In section~4, we make a model independent $Z'$ analysis for LEP~II and
the NLC with all needed radiative corrections.
Mass limits for popular $Z'$ models and for models
with contact interactions are obtained as special cases.
In the case of a positive $Z'$ signal, the interpretation in terms of the
$Z'$ mass or bounds on model parameters is demonstrated.
%
\section{Model Independent $Z'$ Description}
Extra neutral gauge bosons lead to additional neutral current interactions
with SM fermions
\begin{equation}
{\cal L} =
e A_\mu J^\mu_\gamma + g_1 Z_\mu J^\mu_Z + g_2 Z'_\mu J^\mu_{Z'}.
\end{equation}
We assume the gauge group $SU(2)_L\times U(1)_Y\times U'(1)$ at low
energies, where $SU(2)_L\times U(1)_Y$ belongs to the SM and $U'(1)$ to the
$Z'$.
The amplitude of fermion pair production induced by the
new interaction of the $Z'$ is
\begin{eqnarray}
\label{link}
{\cal M}(Z')  = \frac{g_2^2}{s-m_{Z'}^2}
\bar{u}_e\gamma_\beta (\gamma_5 a'_e + v'_e ) u_e \,
\bar{u}_f\gamma^\beta (\gamma_5 a'_f + v'_f ) u_f
\hspace{5cm}
\\ \nonumber
=
 -\frac{4 \pi}{s} \left[
\bar{u}_e\gamma_\beta (\gamma_5 a_e^N + v_e^N ) u_e \,
\bar{u}_f\gamma^\beta (\gamma_5 a_f^N + v_f^N ) u_f \right]
\hspace{5cm}
\end{eqnarray}
\begin{eqnarray}
\label{link2}
\mbox{\ with\ }
a_f^N = a'_f \sqrt{\frac{g_2^2}{4 \pi} \frac{s}{m_{Z'}^2-s}}\ ,\
v_f^N = v'_f \sqrt{\frac{g_2^2}{4 \pi} \frac{s}{m_{Z'}^2-s}}\ \
\mbox{and}\ \ m_{Z'}^2 = M_{Z'}^2-i\Gamma_{Z'} M_{Z'}
\end{eqnarray}
at the Born level.
The minus sign reflects the destructive interference of the $Z'$
contributions with the
photon and $Z$ exchange below the $Z'$ peak.
The normalized couplings to leptons $a_l^N$ and $v_l^N$ will be
restricted by future data.
Far away from resonances, the exact formulae (\ref{link2}) for $a_f^N$ and
$v_f^N$ can be approximated,
\begin{equation}
\label{linkapp}
a_f^N \approx  a'_f \sqrt{\frac{g_2^2}{4 \pi}}\frac{\sqrt{s}}{M_{Z'}},\ \ \
v_f^N \approx  v'_f \sqrt{\frac{g_2^2}{4 \pi}}\frac{\sqrt{s}}{M_{Z'}}.
\end{equation}

In this case, the lepton pair production by a $Z'$ can
be described by effective four fermion contact interactions \cite{eichten}
with $\Lambda = M_{Z'}$
\begin{eqnarray}
\label{contact}
{\cal M}_{eell} = \frac{g^2}{\Lambda^2}\left(
  \eta_{LL}\bar{u}_{e,L}\gamma_\beta u_{e,L} \,
           \bar{u}_{l,L}\gamma^\beta u_{l,L}
+ \eta_{RR}\bar{u}_{e,R}\gamma_\beta u_{e,R} \,
           \bar{u}_{l,R}\gamma^\beta u_{l,R}
\right. \\ \nonumber \left.
+ \eta_{RL}\bar{u}_{e,R}\gamma_\beta u_{e,R} \,
           \bar{u}_{l,L}\gamma^\beta u_{l,L}
+ \eta_{LR}\bar{u}_{e,L}\gamma_\beta u_{e,L} \,
           \bar{u}_{l,R}\gamma^\beta u_{l,R}
\right).
\end{eqnarray}
Assuming lepton universality and setting $g^2/(4\pi) = 1$ by tradition, the
constants of contact interactions can be expressed by the $Z'$ lepton couplings
\begin{equation}
\label{linkkont}
\eta_{LL} = (v^N_l - a^N_l)^2,\ \ \eta_{RR} = (v^N_l + a^N_l)^2,\ \
\eta_{RL} = \eta_{LR} = (v^N_l)^2 - (a^N_l)^2 = \pm\sqrt{\eta_{LL}\eta_{RR}}.
\end{equation}
A general $Z'$ analysis is therefore equivalent to an analysis in terms of
contact interactions with any positive $\eta_{LL}$ and $\eta_{RR}$ taking the
appropriate values for $\eta_{LR}$ and $\eta_{RL}$.
Experimental limits for representative cases of contact interactions are
obtained from LEP data \cite{aleph} and for future $e^+e^-$ colliders
\cite{zepp}. A complete analysis of the parameter space of contact
interactions has not been done.
%
\begin{center}
\begin{minipage}[t]{7.8cm} {
\begin{center}
\ \vspace*{0.5cm}\\
\hspace{-1.7cm}
\mbox{
\epsfysize=7.0cm
\epsffile[0 0 500 500]{coup.ps}
}
\end{center}
}\end{minipage}
\end{center}
\noindent
{\small\bf Fig.~1: }{\small\it
The normalized couplings $(a_l^N,v_l^N)$ of the $Z'$ to charged leptons for
popular $Z'$ models with $M_{Z'} = 4 \sqrt{s}$.
The ranges of the $E_6$ and LR model are indicated as well as the special cases
 $\psi,\ \chi,\ \eta $ and LR.
The abbreviations AA, LL, VV, and RR correspond to contact interactions
\cite{zepp} with $\Lambda = 100 \sqrt{s}$.
The numbers indicate different $Z_\chi$ masses in units of $\sqrt{s}$.
}
\vspace{0.5cm}\\
%
Fig.~1 shows some popular $Z'$ models in terms of the generalized couplings
$a_l^N$ and $v_l^N$.
The $E_6$ GUT and the LRM
correspond to lines parametrized by $\cos\beta$ \cite{e6}
and $\alpha_{LR}$ \cite{lr}
\begin{equation}
\label{zeparm}
J^ \mu_{Z'} = J^\mu_\chi \cos\beta + J^\mu_\psi \sin\beta,\ \ \
J^\mu_{Z'} = \alpha_{LR} J^\mu_{3R} - \frac{1}{2\alpha_{LR}} J^\mu_{B-L}.
\end{equation}
Some completely specified cases are marked by $+\hspace{-2.6mm}\bullet$.
The SSM, where the $Z'$ boson has the same couplings as the SM $Z$ boson, is
also shown for comparison.
Different $Z'$ masses correspond to different points on a straight line in the
$(a_l^N,\ v_l^N)$ plane.
A measurement of the $Z'$ mass at hadron colliders would transform Fig.~1
into constraints to the absolute values of the coupling constants.

%
\section{Model Independent Analysis at the Born level}
At an $e^+e^-$ collider, the following leptonic observables can be
measured \cite{lepobs}
\begin{equation}
\label{obs}
\sigma_t,\ \ A_{FB},\ \ A_{pol},\ \ A_{pol}^{FB}\ \,A_{LR},
\end{equation}
where $\ \ A_{pol},\ \ A_{pol}^{FB}\ \ $ are the polarization asymmetries of
$\tau$ leptons in the final state and $\ \sigma_t,\ \ A_{FB}\ $ and
$\ A_{LR}\ $ are obtained for the production of charged lepton pairs with the
$t$~channel for Bhabha scattering subtracted.
The comparison of future measurements of these observables with its
SM predictions lead to constraints on $a_l^N$ and $v_l^N$.
The obtained limits are sensitive to the expected experimental
errors and radiative corrections.

As in \cite{zpsari}, we assume a 1\% systematic error due to the luminosity
uncertainty and 0.5\% due to the event selection of leptons.
The systematic error of the forward backward asymmetry is assumed to be
negligible and that of the left right polarization asymmetry to be
$\Delta A_{LR} = 0.3\%$.

As statistical errors for $N$ detected events we take:
\begin{equation}
\label{stat}
\frac{\Delta\sigma_t}{\sigma_t}=\frac{1}{\sqrt{N}},\ \ \
\Delta A_{FB} = \Delta A_{pol} = \Delta A = \sqrt{\frac{1-A^2}{N}},\ \ \
\Delta A_{LR} = \sqrt{ \frac{1 -(P A_{LR})^2}{N P^2} }\ .
\end{equation}
$P$ is the degree of polarization set to one here.
We assume that for 75\% of the $\tau$ events the final state
polarization can be measured \cite{lepobs}. We took an integrated luminosity
$L_{int}=0.5fb^{-1}$ for LEP~II and $L_{int}=20fb^{-1}$ for the NLC.

The systematic and statistical errors have then been combined quadratically.
For completeness, we list the combined errors for every of the considered
observables
\begin{eqnarray}
\label{experr}
\frac{\Delta\sigma_t}{\sigma_t} =1.6\%,\ \Delta A_{FB} = 1.1\%,\
\Delta A_{pol} = \Delta A_{pol}^{FB} = 2.8\%,\
\Delta A_{LR} = 1.4\% ,\ \ \mbox{LEP~II}
\nonumber\\
\frac{\Delta\sigma_t}{\sigma_t} =1.3\%,\ \Delta A_{FB} = 0.5\%,\
\Delta A_{pol} = \Delta A_{pol}^{FB} = 1.2\%,\
\Delta A_{LR} = 0.7\% ,\ \ \ \ \mbox{NLC}.
\end{eqnarray}

In theories with lepton universality, neglecting the difference of lepton
masses, three observables are related at the Born level
\begin{equation}
\label{relation}
A_{LR} =  A_{pol} = \frac{4}{3} A_{pol}^{FB}.
\end{equation}
A violation of these relations would indicate that there exist new physics
beyond the SM which are not due to extra $Z$ bosons.

Under the assumption that both asymmetries have the same experimental errors,
$A_{pol}^{FB}$ gives always worse $Z'$ constraints than $A_{pol}$.
Therefore, it is not considered in the further analysis. Among $A_{LR}$ and
$A_{pol}$, the observable with the smaller experimental error gives better $Z'$
constraints.
$A_{LR}$ has about four times higher event rates being sensitive to
all leptons. However, equation (\ref{stat}) shows that
its predictive power is reduced by degrees of polarization smaller than 100\%.
With degrees of polarizations of 50\% or less, measurements of
$A_{LR}$ add no new information to the $Z'$ search.
Therefore, an upgrade of the
luminosity by a factor 2 to 3 should be preferred compared to polarized beams
because it reduces the error of all observables.
In case of a non-zero $Z'$ signal, quark pair production by polarized beams
would help to determine the $Z'$-quark couplings as far as the final quark
polarization is not measurable.

We now demonstrate the model independent $Z'$ analysis at
the Born level. The result are approximate analytical formulae making
transparent how the model independent $Z'$ constraints arise from different
observables.

The observable $\sigma_t$ can ``see'' a signal of a $Z'$, if
\begin{equation}
\label{chi2r}
\chi^2=\left(\frac{\Delta^{Z'}\sigma_t}{\Delta\sigma_t}\right)^2 \ge F_\chi^2.
\end{equation}
$\Delta\sigma_t$ is the combined experimental error as given in
(\ref{experr}), $F_\chi^2$ is a number depending on the confidence level and
the details of the analysis and $\Delta^{Z'}\sigma_t$ is the deviation of
$\sigma_t$ from its SM prediction $\sigma_t^{SM}$ due to a $Z'$.

In this section, we will neglect all
imaginary parts of the $Z$ and $Z'$ propagators,
\begin{equation}
\label{chiz}
\chi_Z =\frac{\sqrt{2}G_\mu M_Z^2}{4\pi\alpha} \frac{s}{s-M_Z^2+iM_Z\Gamma_Z}
\approx \frac{\sqrt{2}G_\mu M_Z^2}{4\pi\alpha} \frac{s}{s-M_Z^2}.
\end{equation}
Neglecting the quadratic $Z'$ contributions and taking into account
that the leptonic vector coupling of the SM $Z$ boson $v_l$ is small against
1 and the axial vector coupling $a_l=-1/2$, one obtains
\begin{equation}
\label{deltazp}
\Delta^{Z'}\sigma_t \approx \frac{4\pi\alpha}{3s}
2 \left[ (v_l^N)^2 + \chi_Z a_l^2 (a_l^N)^2 \right].
\end{equation}
The first contribution is due to the $\gamma Z'$ interference, the second is
due to the $Z Z'$ interference. Equation (\ref{deltazp}) together with
(\ref{chi2r}) give the constraint on $a_l^N$ and $v_l^N$ by $\sigma_t$,
\begin{equation}
\label{sigc}
\left(\frac{v_l^N}{H_t}\right)^2 + \left(\frac{a_l^N}{H_t}\right)^2
\frac{\chi_Z}{4} \ge 1,\ \ \
H_t = \sqrt{\frac{F_\chi\alpha}{2}\frac{\sigma_t^{SM}}{\sigma_t^{QED}}
\frac{\Delta\sigma_t}{\sigma_t^{SM}} }.
\end{equation}
In (\ref{sigc}), $\sigma_t^{QED}=\frac{4\pi\alpha^2}{3s}$ is the QED cross
section.
We conclude from (\ref{sigc}) that the observable $\sigma_t$ cannot exclude an
ellipse of ($a_l^N$,\ $v_l^N$) around the point (0,0).

Similar considerations can be done for the observables $A_{FB}$ and $A_{LR}$.
They lead to the following exclusion regions
\begin{equation}
\label{afbc}
\left|
\left(\frac{v_l^N}{H_{FB}}\right)^2 - \left(\frac{a_l^N}{H_{FB}}\right)^2
\frac{(3 - A_{FB}^{SM} \chi_Z)
\frac{1}{4}}{A_{FB}^{SM}-\frac{3}{16}\chi_Z} \right| \ge 1,\ \ \,
H_{FB}=\sqrt{\frac{F_\chi\alpha}{2}\frac{\sigma_t^{SM}}{\sigma_t^{QED}}
\Delta A_{FB} } \left[ A_{FB} -\frac{3}{16}\chi_Z\right] ^{-1/2},
\end{equation}
%
\begin{equation}
\label{alrc}
\left|
\left(\frac{v_l^N}{H_{LR}}\right)\left(\frac{a_l^N}{H_{LR}}\right)
\right|\ge 1,\ \ \
H_{LR}=\sqrt{\frac{F_\chi\alpha}{2}\frac{\sigma_t^{SM}}{\sigma_t^{QED}}
\Delta A_{LR} } \left[ 1 + \frac{1}{4}\chi_Z\right] ^{-1/2}.
\end{equation}
In the last two equations, $A_{FB}^{SM}$ and $A_{LR}^{SM}$ are the SM
predictions for the corresponding observables. The two considered asymmetries
$A_{FB}$ and $A_{LR}$ cannot
exclude the area between hyperbolas. We see that the three considered
observables are complementary in excluding $Z'$ contributions. The equations
(\ref{sigc}) - (\ref{alrc}) reflect all qualitative features of the results
with radiative corrections to be discussed in the next section.

$H_t,\ H_{FB}$ and $H_{LR}$ are a measure of the sensitivity to $Z'$ effects.
They are all proportional to the square rout of the experimental
error. Taking into account only the statistical error, we
obtain a simple scaling law of the $Z'$ exclusion limits with the
integrated luminosity $L_{int}$ and the c.m. energy $\sqrt{s}$,
\begin{equation}
\label{scale}
a_l^N,\ v_l^N \sim H_t,\ H_{FB},\ H_{LR} \sim N^{-1/4} \sim
\left( L_{int}/s\right)^{-1/4}.
\end{equation}
Together with (\ref{linkapp}), we get for the scaling of the $Z'$ mass limit
\begin{equation}
\label{scalem}
M_{Z'}^{lim} \sim \sqrt{s}/a_l^N,\ \sqrt{s}/v_l^N \sim
\left( L_{int} s \right)^{1/4}.
\end{equation}
This agrees with the results of \cite{zepp} and \cite{rizzo}.

In case of a positive $Z'$ signal $(\bar{a}_l^N,\ \bar{v}_l^N)\ne (0,0)$,
experimental constraints on ($a_l^N,\ v_l^N$) could be obtained by
modifications of the above
deviation.  In contrast to the previous discussion, the size of the area in the
$(a_l^N,\ v_l^N)$ plane which cannot be resolved is now proportional to
$M_{Z'}$ and  centered around the point $(\bar{a}_l^N,\ \bar{v}_l^N)$.
Suppose, that the $Z'$ couplings are proportional to some model parameter
$\alpha_{Z'}$ and that the $Z'$ mass is $M_{Z'}$. Suppose further, that
$\alpha_{Z'}$ and $M_{Z'}$ can be measured with errors $\Delta\alpha_{Z'}$ and
$\Delta M_{Z'}$. Then, the scaling of these errors with the $Z'$ mass is
\begin{equation}
\label{scaleb}
M_{Z'}\rightarrow c\,M_{Z'}\; \Longrightarrow\;
\Delta M_{Z'}\rightarrow c^3\,\Delta M_{Z'}\; \mbox{ and }\;
\Delta \alpha_{Z'}\rightarrow c^2\,\Delta\alpha_{Z'}.
\end{equation}
The scaling law discussed above is only valid for a $Z'$ mass larger than
the c.m. energy and smaller than $M_{Z'}^{lim}$. Equation (\ref{scaleb})
agrees well with
our numerical results and those of \cite{zpsari} and \cite{rizzo}.

The squares of the couplings $(a_l^N)^2,\ (v_l^N)^2$ are
expected to have a gaussian error because
they are proportional to parts of cross sections. These squares don't feel
the sign of $a_l^N$ as well as the equations (\ref{sigc}) and (\ref{afbc}).
However, for a non-zero $Z'$ signal, the observables $A_{LR},\ A_{pol}$ and
$A_{pol}^{FB}$ lead to deviations from their SM predictions which have a
different sign for different signs of $a_l^N$. Furthermore,
the exact deviation of  (\ref{sigc}) - (\ref{alrc})
shows also a weak dependence on the sign of $a_l^N$ due to
terms of the order $O(v_l^N/a_l^N)$ neglected
in the approach above. Hence, in a model independent $Z'$ search, the simple
couplings and not its squares should be analyzed although their errors are
not distributed gaussian.

%
\section{The Analysis including Radiative Corrections}
All calculations are done using the code {\tt ZCAMEL} \cite{zcamel},
which contains the $O(\alpha)$ QED corrections with soft
photon exponentiation for the process
$e^+e^- \rightarrow (\gamma,\ Z,\ Z') \rightarrow f^+f^-$.
A cut $\Delta$ on the energy of the radiated photon
$\Delta = 0.7 >E_\gamma/E_{beam}$ has been applied to remove the radiative
tail and the potentially dangerous background \cite{zpsari}.
The analytical formulae of the QED corrections used here are given in
\cite{zpsigafb}.
The SM electroweak corrections must not be considered because they cancel in
the {\it deviation} of an observable from its SM value.
The main effect of the radiative corrections is a relaxation of the
$Z'$ limits of roughly 10\% compared to the Born result.

The constraints to $a_l^N$ and $v_l^N$ are shown for LEP~II in Fig.~2a and
for the NLC in Fig.~2b.
All (one sided) exclusion limits are at the 95\% c.l., which corresponds to
$\chi^2=F_\chi^2=2.69$ for one degree of freedom \cite{kroha}.
They should be compared with the approximate analytical results of
 (\ref{sigc}) - (\ref{alrc}).
In general, the constraints to $a_l^N$ and $v_l^N$ are worse at LEP~II
because it has larger errors compared to the NLC, see (\ref{experr}).
\ \vspace{1cm}\\
\begin{minipage}[t]{7.8cm}{
\begin{center}
\hspace{-1.7cm}
\mbox{
\epsfysize=7.0cm
\epsffile[0 0 500 500]{zpmilep.ps}
}
\end{center}
\noindent
{\small\bf Fig.~2a: }{\small\it
The areas in the $(a_l^N,\ v_l^N)$ plane excluded by different observables
at LEP~II with 95\% confidence. The allowed range always contains the point
$(0,0)$.
}
}\end{minipage}
\hspace{0.5cm}
\begin{minipage}[t]{7.8cm} {
\begin{center}
\hspace{-1.7cm}
\mbox{
\epsfysize=7.0cm
\epsffile[0 0 500 500]{zpminlc.ps}
}
\end{center}
\noindent
{\small\bf Fig.~2b: }{\small\it
The same as Fig.~2a but for the NLC. The numbers
indicate the $Z_\chi$ mass in units of $\sqrt{s}$.
}
}\end{minipage}
\vspace*{0.5cm}
%

The Figures 2a and 2b are model independent. They can be used to obtain
mass limits for particular $Z'$ models.
Consider the $Z_\chi$ arising in the breaking chain of
the $E_6$ GUT \cite{e6} at Fig.~2b.  We see that $\sigma_t$ and $A_{LR}$
give mass limits of $M_{Z'} > 2.2\,TeV$ at the NLC, while the constraint from
$A_{FB}$ is $M_{Z'} > 1\,TeV$. The corresponding intersection point is
outside of Fig.~2b.
This is in agreement with the mass limits
obtained for the $E_6$ GUT under the same experimental conditions in
\cite{zpsari}.  The mass limits for any other $Z'$ model from LEP~II and the
NLC could be obtained from Fig.~2a and Fig.~2b in a similar way. The limits
on four fermion contact interactions agree with \cite{zepp} if one takes into
account the differences between the two analyses.

For $M_{Z'}\approx \frac{2}{3}M_{Z'}^{lim}$, any point $(a_l^N,\ v_l^N)$ is
covered by at least two observables, where $M_{Z'}^{lim}$ is the
discovery limit for the same $Z'$ model, see Figs.~2.
The leptonic couplings of the $Z'$ can then be measured as
demonstrated in Figs.~3. Note the difference in the scale compared to Figs.~1
and 2.
The area of $(a_l^N,\ v_l^N)$ allowed with 95\% confidence by one observable
alone is defined by $\chi^2=4$.
The combined allowed area around the $+\hspace{-2.6mm}\bullet$
of all three observables is obtained taking $\chi^2=7.7$.
A more concrete interpretation of this area demands specifications of
$M_{Z'}$ or the $Z'$ model. Fixing $Z'=Z_\chi$, Fig.~3b gives the
errors of the measurement of $M_{Z'}$. Fixing $M_{Z'}=3\sqrt{s}=1.5\,TeV$,
Fig.~3b gives the errors of the measurement of model parameters as for
$\cos\beta$ in the $E_6$ model, $\alpha_{LR}$ in the LRM or areas of
confusion between these models.
\ \vspace{1cm}\\
\begin{minipage}[t]{7.8cm}{
\begin{center}
\hspace{-1.7cm}
\mbox{
\epsfysize=7.0cm
\epsffile[0 0 500 500]{nlcdef.ps}
}
\end{center}
\noindent
{\small\bf Fig.~3a: }{\small\it
The area ($a_l^N,\ v_l^N$) allowed by $\sigma_t,\ A_{FB}$ and $A_{LR}$
separately with 95\% confidence. $Z'=Z_\chi$ with $M(Z_\chi) = 3 \sqrt{s}$
was chosen and marked by $+\hspace{-2.5mm}\bullet$. The combined area allowed
by all three observables is indicated by the thick dotted line.
}
}\end{minipage}
\hspace{0.5cm}
\begin{minipage}[t]{7.8cm} {
\begin{center}
\hspace{-1.7cm}
\mbox{
\epsfysize=7.0cm
\epsffile[0 0 500 500]{coupdef.ps}
}
\end{center}
\noindent
{\small\bf Fig.~3b: }{\small\it
Demonstration of the measurement of the $Z'$ mass and its couplings to
charged leptons at the NLC.
The thick dotted line around the the 3 is copied from
Fig.~3a. The thin dotted lines around the 2 (4) correspond to $M_{Z'}=2\
(4)\sqrt{s}$.  For all other lines see Fig.~1.
}
}\end{minipage}
\vspace*{0.5cm}
%

If future measurements are inconsistent with the existence of a $Z'$, they
should be described in the more general scheme of four fermion contact
interactions.  In contrast to $Z'$ models, negative
signs of $\eta_{LL}$ and $\eta_{RR}$ are allowed and $\eta_{LR}$ and
$\eta_{RL}$ are independent of $\eta_{LL}$ and $\eta_{RR}$.  An inconsistency
of measurements with $Z'$ theories could arise due to the violation of
(\ref{relation}) or due to a zero allowed area for ($a_l^N,v_l^N$) after
combining all observables.

Consider the case of a very weakly coupling $Z'$.
The resonance curve of such a $Z'$ is very narrow and high.
It can be detected only very near its resonance.
The observables near a resonance cannot be described by four fermion contact
interactions.  Radiative corrections become much more
important making the Figs.~2 and 3 dependent on the $Z'$ mass.
However, we have proven that even for the extreme case $M_{Z'}=505\,
GeV=1.01\sqrt{s}$ at the NLC, the effect of radiative corrections is only a
relaxation of the limits on $a_l^N$ and $v_l^N$ by 30\% leaving Figs.~2b and
3 qualitatively unchanged.  Eventually, a weakly coupling $Z'$ would never
be missed at energies above the $Z'$ peak, because its radiative tail
is proportional to $M_{Z'}/\Gamma_{Z'}$.

To summarize, a model independent $Z'$ analysis is made for future $e^+e^-$
colliders. Numerical examples are given for LEP~II and a
500\,GeV $e^+e^-$ collider. Assuming lepton universality, all available
leptonic observables except Bhabha scattering are considered.
Approximate analytical formulae for $Z'$ exclusion limits
and scaling laws are obtained.
All needed radiative corrections are included in the analysis and their
effects on $Z'$ constraints are discussed.
It is shown, how the different observables give complementary constraints to
$Z'$ physics.
Polarized beams add useful information to the $Z'$ search only for degrees of
polarization essentially larger than 50\%.
In case of a positive $Z'$ signal,
it is demonstrated, under which conditions the $Z'$ couplings to leptons
and the $Z'$ mass can be measured.
Connections to four fermion contact interactions are shown as well as to
popular $Z'$ models.

%
%
\ \vspace{1cm}\\ {\Large\bf Acknowledgement\hfill}\vspace{0.5cm}\\
I would like to thank D. Zeppenfeld for useful discussions and T. Riemann for
the careful reading of the manuscript.
Further, I would like to thank the University of Wisconsin for the
kind hospitality in Madison, where a part of this work was performed.
%

\end{document}